\documentclass[aps,pre,preprint,groupedaddress,showpacs]{revtex4}
\usepackage{epsfig}
\begin{document}

\preprint{}

\title{Turbulence near cyclic fold bifurcations in birhythmic media}

\author{Dorjsuren Battogtokh} 
\email{dbattogt@vt.edu}
\author{John J. Tyson}
\affiliation{
Department of Biology, Virginia Polytechnic and State University
Blacksburg, VA 24061-0106
} 

\date{\today}

\begin{abstract}
We show that at the onset of a cyclic fold bifurcation,
a birhythmic  medium composed of glycolytic oscillators 
displays turbulent dynamics. By computing the largest Lyapunov 
exponent, the spatial correlation function,  and the average 
transient lifetime, we classify it as a weak turbulence 
with transient nature. Virtual heterogeneities
generating  unstable fast oscillations 
are the  mechanism of the transient turbulence.  
In the presence of  wavenumber instability, unstable oscillations 
can be reinjected leading to 
stationary turbulence. We also find similar turbulence
in a cell cycle model. These findings  
suggest that  weak turbulence may be universal 
in  biochemical birhythmic media exhibiting cyclic 
fold bifurcations.
\end{abstract}

\maketitle
\section{Introduction}
In studies of chemical turbulence
 in reaction diffusion systems near 
a Hopf bifurcation, a reduction of the model to the complex Ginzburg Landau 
equation(CGLE) is very useful \cite{kura1,cross}. First, it allows to 
define a parameter set in the model leading to turbulence  without 
carrying out simulations 
\cite{kbn}. Secondly, the
detailed knowledge of the CGLE's dynamics can be very  helpful \cite{
nozaki,saka,shraiman,kramer1}, because  mathematical models from different 
disciplines displaying dynamics near a Hopf bifurcation
obey the same qualitative dynamics of the CGLE \cite{kura2}. 

However, the CGLE alone is insufficient 
for a qualitative description of realistic models in a neighborhood of 
a Hopf bifurcation, when other bifurcations occur nearby
\cite{kramer2,show1}. For example, near a 
supercritical Hopf bifurcation point, another stable limit cycle may exist, 
so that, depending on initial conditions, oscillations with two different 
frequencies and amplitudes are possible. Such a situation called birhythmicity
is a characteristic feature of a number of well known models in biochemical 
oscillations \cite{gold1,borisuk}. For these systems, the CGLE cannot be used 
without appropriate modifications. Often, the best way to approach 
these problems is simulations of the original models \cite{show1,yama}.

To the best of our knowledge, little is known about turbulence in birhythmic 
media. Intuitively, in a regime of a strong wavenumber instability, 
birhythmicity should not be a factor. Therefore, turbulence in 
homogeneous birhythmic media and in coupled limit cycle oscillators
should have similar characteristics. In the absence of wavenumber instability,
high frequency oscillations are supposed to suppress slow 
oscillations and restore uniform oscillations. 
But at the onset of a cyclic fold (CF) bifurcation in birhythmic media of 
a biochemical origin, high frequency oscillations may be  unstable. 
Thus, a complete suppression of slow oscillations may not be achieved in
these systems. On the contrary, if unstable oscillations emerge persistently, 
complex spatiotemporal motions are possible.

The goal of this work is to  show that near cyclic fold bifurcations
in birhythmic media, virtual heterogeneities creating unstable oscillations 
can lead to a peculiar turbulence, 
intermittency of small and large
amplitude oscillations.
We will first compute complex
spatiotemporal behavior in a birhythmic 
medium composed of glycolytic oscillators. By calculating 
the maximal Lyapunov exponent, the spatial correlation function, and the 
average transient lifetime, we will provide evidence 
that this behavior is weak transient turbulence. In the presence of 
wavenumber instability 
transient turbulence may become stationary. Mathematically, the instability of the
faster oscillations is a result of  a CF 
bifurcation driven by the terms representing enzymatic regulations,
suggesting that weak turbulence may be common 
in biochemical birhythmic media exhibiting CF bifurcations.
As further evidence, we demonstrate weak turbulence 
in a cell cycle model. 
A biological system where weak turbulence might possibly  be found 
is presented in the closing section.

\section{A birhythmic medium of glycolytic oscillators}
Let us introduce a birythmic medium composed of glycolytic oscillators,

\begin{eqnarray}
\frac{d\alpha}{dt}=\nu+\frac{\sigma_i \gamma^n}{K^n+\gamma^n}-\sigma \phi
+D_{\alpha}\Delta\alpha,   {\hspace{0.62cm}}\\
\frac{d\gamma}{dt}=Q \sigma \phi -k_s \gamma 
-\frac{Q \sigma_i \gamma^n}{K^n+\gamma^n}+D_{\gamma}\Delta\gamma,\\
\phi=\frac{\alpha (1+\alpha) (1+\gamma)^2}{L+ (1+\alpha)^2  (1+\gamma)^2}.
\hspace{2.15cm}\nonumber
\end{eqnarray}

In Eqn. (1-2), $\alpha$ and $\gamma$ represent dimensionless substrate
and product concentrations of glycolytic reactions, $K$,
$n$, $\nu$, $\sigma_i$, $\sigma$, $k_s$,
$L$ and $Q$ are parameters. For convenience, we assume $Q\equiv 1$ throughout
this paper. $D_{\alpha}$, $D_{\gamma}$ are  
diffusion constants for the substrate and product. 
Our units of time and space are {\rm sec} and {\rm cm}, respectfully. 
When $D_{\alpha}=0$, $D_{\gamma}=0$ and $\sigma_i=0$, 
Eqn. (1-2) is called the glycolytic oscillator. The term
$\frac{\sigma_i \gamma^n}{K^n+\gamma^n}$ represents
substrate recycling that drives  birhythmicity. 
Recently,  in Ref. \cite{gold2}, Eqn. (1-2)  were shown to support 
multiple wave fronts.  Our concern in this paper is 
a different parameter region where irregular spatiotemporal  
motions develop. 

A phase plane analysis of Eqn. (1-2) shows that the mechanism of birhythmicity 
is two regions of negative slope in the product nullcline \cite{gold1}. 
A convenient way  to illustrate birhythmicity is a bifurcation diagram. 
We used a  well known software package, AUTO \cite{doedel}, for 
bifurcation analysis of the local model($D_{\alpha}=D_{\gamma}=0$
in Eqn. (1-2)).  Solid lines in Fig. 1 show stable steady states,
dashed lines show unstable steady states.  Stable limit
cycles are shown by filled symbols, unstable
limit cycles by open circles. Filled circles represent
small amplitude oscillations with high frequencies. 
Large amplitude oscillations with low frequencies are shown by 
filled diamonds in Fig. 1. A Hopf bifurcation point $HB$ is located
at $\sigma_{i,cr} \approx 1.282$.  There are two 
CF bifurcations in Fig. 1, where 
stable limit cycles are replaced
by unstable ones. Between these two CF points,
which occur  at 
$\sigma_{i,CF^1} \approx 1.77$ and $\sigma_{i,CF^2} \approx 1.83$,
two stable limit cycles coexist. Therefore, 
depending on initial conditions, one of the 
limit cycles will be selected in simulations of 
the glycolytic oscillator with substrate recycling. 

A general mechanism of turbulence in oscillatory
reaction diffusion systems is wavenumber instability, i.e.,
instability of uniform oscillations against phase-like 
fluctuations \cite{kura1}. 
In Eqn. (1-2), there are
two different uniform oscillations that might undergo 
wavenumber  instability. We want to provide evidence that
these oscillations are stable against phase-like fluctuations
for the parameters in Fig. 1.
For the fast, uniform oscillations
which originate from the Hopf bifurcation point 
shown by filled circles in Fig. 1,  the stability condition 
can be obtained by reducing Eqn. (1-2) to the CGLE,
\begin{equation}
\dot A=(1+ i c_0) A-(1+i c_2)|A|^2 A+(1+i c_1) \Delta A.
\end{equation}
In Eqn. (3) $A$ is the complex amplitude, and $\omega$, $\beta$, $\gamma$ are
real parameters. The CGLE has a uniform oscillatory 
 solution,  $A=exp(i (c_0-c_2) t)$, that is stable if the  condition
$ 1+ c_1 c_2 > 0$ holds.
In the appendix, we calculated $c_0$, $c_1$ and 
$c_2$ corresponding to Eqn. (1-2).
Our results show that the uniform  oscillations are stable for
the parameters used in Fig. 1, for any $D_{\alpha}>0$
and  $D_{\gamma}>0$. For $D_{\alpha}=D_{\gamma}$ we find that  $c_1=0$. 
Hence, the parameter region we are interested in is deep inside  
the Benjamin-Feir stability region  given by  $ 1+ c_1 c_2 > 0$. 
Although  the CGLE is valid only near the $HB$ point, 
it is likely that the uniform  oscillations will remain stable 
until the the next bifurcation in the system, i.e., 
$CF^1$ in Fig. 1 \cite{kura2}. Next,  consider the uniform 
oscillations with low frequencies.
Unlike the case of fast oscillations, 
no analytic approach is available in this case.
Note that oscillations shown by filled circles and diamonds in Fig. 1
occur at the same parameters. Therefore, it is rather unlikely 
that the slow oscillations undergo wavenumber instability, 
contrary to the fast ones. Thus, we can assume that uniform, 
slow oscillations are also stable.

It is known that strong perturbations can switch oscillations from one stable 
orbit to another in the glycolyctic model with substrate recycling \cite{gold1}. 
Therefore, even if both uniform oscillations in Eqn. (1-2)  are stable 
against wavenumber instability, strong perturbations can excite the system 
by switching the oscillations. This kind of excitability, however, will 
not lead to turbulence; in the parameter 
interval $[\sigma_{i,CF^1},\sigma_{i,CF^2}]$,
the fast oscillations will suppress 
the slow ones as time progresses. 
But for $|\sigma_i-\sigma_{i,CF^1}| \ll 1 $
where the fast  oscillations become unstable, it is apparent  that
a complete suppression of slow oscillations is impossible. 
Here, because of complex interactions between stable, slow and  unstable, 
fast oscillations,  interesting spatiotemporal dynamics might develop.
Therefore, we carried out a detailed numerical study in the 
neighborhood of $CF^1$.

\section{Weak turbulence 
in a birhythmic medium of glycolytic oscillators}
For numerical integrations of Eqn. (1-2) in 
one spatial dimension, we used the fourth order
Runge-Kutta method. Diffusion terms were approximated by the 
finite difference method. Numerical parameters are $\delta x=0.005{\rm cm}$, 
$\delta t=0.05{\rm s}$. The system size is defined as $l=N \delta x$, 
where $N$ is the number of spatial grid points. 
In this paper we present results for
periodic boundary conditions, but  we also tested the main results with 
no-flux boundary conditions. We also tested selected examples with smaller
values of $\delta x$ and $\delta t$ for fixed $l$.
Our simulations show that 
Eqn. (1-2) are sensitive to initial conditions. By choosing 
initial conditions as small perturbations of uniform, 
slow oscillations with large amplitudes, 
we found that these oscillations are stable
for $\sigma_i<\sigma_{i,CF^2}$. But, near and to the left of $CF^1$, 
uniform, fast  oscillations with small amplitudes
are found to be unstable. They spiral out from unstable  orbits towards
the  orbit of stable, large amplitude oscillations.  
For strong perturbations near the $CF^1$ bifurcation point,
we found spatiotemporal irregular motions in Eqn. (1-2). 

Fig. 2 shows a gray scale plot of spatiotemporal 
dynamics in Eqn. (1-2). Oscillations between the white and black colors show 
large amplitude oscillations displayed by $\gamma(x,t)$. There are also 
oscillations with  higher frequency and smaller amplitude in Fig. 2.
Because the latter ones  are unstable, they  
can not suppress large amplitude domains.
Although uniform, large amplitude oscillations are stable 
against small fluctuations,
phase slips created by strong initial perturbations cannot be 
eliminated as time progresses. As a result, spatially 
nonuniform distributions of concentrations are seen at 
given time moments, Fig. 3. On the phase plane, these 
nonuniform distributions generate motions attracted by unstable 
orbits around the inner cycle shown in Fig. 4. We found that such 
unstable orbits act as a weak, virtual heterogeneity emerging
randomly. They cannot entrain the bulk oscillations, 
but in their presence, phase slips cannot be eliminated. 
Instead, persistent spatiotemporal irregular motions develop.

To characterize the irregular motions in Fig. 2,
we calculated the maximum Lyapunov exponent $\lambda^{max}_{lyap}$
in $2N$ dimensional phase space \cite{wolf}. First we made a very 
long run of Eqn. (1-2) to confirm that the turbulence is stationary. 
Then, by using the same initial conditions, 
we simulated Eqn. (1-2) and its linear
system for  computation of $\lambda^{max}_{lyap}$
for $T_1=2 \cdot 10^5{\rm s}$. 
We found that the largest Lyaponov exponent 
is positive and small,
$\lambda^{max}_{lyap} \approx 2\cdot 10^{-3}$. We also calculated 
a two-point correlation function, $C(x)=<\gamma(x_0,t) \gamma(x_0+x,t)>$, where
$<..>$ stands for an average over space and time \cite{bataa}. 
Fig. 5 shows that 
$C(x) \approx const$ at small values of $x$, indicating 
strong local coupling 
and an absence of short waves. A power-law decay of the correlation function
at intermediate values of $x$ implies the  presence of chaotic  
motions. We found that the slope is $\kappa \approx -0.15$. 
We also found no significant variations of $\kappa$
and $\lambda^{max}_{lyap}$ 
with changes of $\sigma_i$ and $l$. The small values of $\kappa$
and $\lambda^{max}_{lyap}>0$ suggest that spatiotemporal 
irregular motions shown in Fig. 2-4 can be characterized 
as a weak turbulence. 

We found that in Eqn. (1-2), stationary irregular motions  
can develop only for certain initial conditions and 
system sizes. In simulations with 
different initial conditions and system sizes, we observed
sudden collapses  of turbulent dynamics. 
Collapse of turbulence in Eqn. (1-2) means a complete suppression
of small amplitude oscillations.
Thus, we defined the transient lifetime of turbulence $t_p$ as the
time interval from initial conditions to the moment 
when all oscillators come within a distance $d$ of the orbit of stable, slow, 
large amplitude oscillations. In our simulations we used $d=0.03$.
Following Ref.  \cite{show2}, we plot an average transient 
lifetime $t_p$ versus the system size $l$ in Fig. 6. 
Here, each filled circles is an  average of 20 simulations  
with different initial conditions. Fig. 6 shows that, 
as the system size increase, $t_p$ grows exponentially.

For some initial conditions, when $l$ is close to 
$2{ \rm cm}$,
the turbulent solution does not collapse. 
The inset in Fig. 6 shows the number of cases, among 20 different 
simulations, when a collapse of turbulence has not occurred by
$T=10^6{\rm s}$. (These cases were not included in calculations of 
the filled  circles in Fig. 6.) We continued two cases
in the inset (at $l=1.75{\rm cm}$) up to $T=10^8 {\rm s}$ and did
not observe a collapse of motions near the inner cycle in Fig. 4. 

Numerical experiments indicate that if  virtual heterogeneities 
reside sufficiently far from 
each other, a stationary pattern is possible
in the interval 
$\sigma_i \in [1.055,1.075]$.  Fig. 7 gives an example
of such a pattern. Here virtual heterogeneities are 
located from each other by distances between $ 0.5{\rm cm}$ and  
$1{\rm cm}$. Note that these quasi-periodic structures
are not related to a  Turing
instability, which emerges due to differences in diffusion
coefficients. Unstable oscillations at the onset of a CF  bifurcation
are the instabilities leading to these structures. 
The cellular structures in Fig. 7 are breathing
because of the unstable oscillations. Numerical results show 
that as  the system size increases, the cells breath coherently.

A collapse of turbulence can be prevented if there is a reinjection
mechanism for the unstable oscillations generated by the virtual 
heterogeneities. Naturally, wavenumber instability can be such
a  mechanism. Using our calculations in the appendix, 
we simulated Eqn. (1-2) for
parameters when the corresponding CGLE displays a wavenumber instability and 
found that small amplitude oscillations exhibit phase instability
near Hopf bifurcation. For $|\sigma-\sigma_{i,CF^1}| \ll 1$ we found 
stationary weak turbulence.

\section{weak turbulence in a cell cycle model}
In section III, we demonstrated that the $CF^1$ bifurcation point
is crucial  for turbulence in  Eqn. (1-2). 
Mathematically, the term representing  substrate recycling drives
CF bifurcations. In models of biochemical oscillations,
terms representing enzymatic activities naturally 
arise. As an enzyme can quickly switch from being 
active to inactive and back again, 
ideal conditions for CF bifurcations exist 
in these models. Therefore, other
biochemical reaction diffusion models may also display
the weak turbulence discussed in the previous section.
As an example, consider  a three variable model of the 
budding yeast cell cycle,

\begin{widetext}
\begin{eqnarray}
%%%Equation#1
\frac{d X}{dt} =m (k_1+k_2 T) - ( k_3+k_4 Y
+k_5 Z)X + D_X\Delta X, {\hspace{3cm}}\\
%%%Equation#2
\frac{d Y}{dt}=\frac{(k_6+k_7 Z)(1-Y)}{J_1+1-Y}-
\frac{(k_8 m+k_9 X) Y}{J_1+Y} +  D_Y\Delta Y,  
{\hspace{2.75cm}}\\
%%%Equation#3
\frac{d Z}{dt}=(k_{10}+k_{11} X) -k_{12} Z +  D_Z\Delta Z,
{\hspace{5.8cm}}\\
%%%Equation#4
T=G(X,P, J_2, J_2),{\hspace{8.7cm}}\\
G(a,b,c,d)= 
\frac{2 a d}{ b-a+b c +a d +\sqrt{(b-a+b c +a d)^2-4 a d (b-a)}},
 \hspace{0.45cm}
\end{eqnarray}
\end{widetext}
where the transcription factor $T$  for $X$ is given by the 
Goldbeter-Koshland function $G$ \cite{gold1}. $X$, $Y$, $Z$ 
are dimensionless variables 
and $m$ is a dimensionless parameter.  Our units of time and space are 
{\rm min} and {\rm cm}, respectfully. 

When $D_X=D_Y=D_Z=0$, Eqn. (4-8) are  the reduced version of 
a budding yeast cell cycle model \cite{kathy}.
Here, $X$ represents the  concentration of  cyclin-dependent 
protein kinase(CDK),
$Y$ and $Z$ are concentrations of 
two different anaphase promoting complexes(APC),
APC/Cdh1 and APC/Cdc20 respectively. 
In Eqn. (4-5), $m$ represents the cell's  mass, which will be used as a primary
bifurcation parameter.  Eqn. (4-8) display CF bifurcations 
as shown in Fig. 8.
For small $m$, Eqn. (4-8) also display saddle node bifurcations, 
a feature universal in cell cycle
models \cite{borisuk,kathy}. Here, our concern is 
the neighborhood of $CF^1$  in Fig. 8. 

There are no experimental measurements of  diffusion coefficients for 
CDK, and APC factors. But it is known that diffusion coefficients
of average sized proteins in cytoplasm are order of 
$10^{-4}\frac{{\rm cm^2}}{{\rm min}}$ or smaller \cite{woj,zubay}. 
As our goal is a demonstration of weak turbulence 
in a representative model of biochemical oscillations, 
we choose $D_X$ and $D_Z$ arbitrarily subject to this upper bond. For
simplicity, we assume  $D_Z=0$. 

For simulations of Eqn. (4-8) we used the same method as in the previous
section with $\delta t=0.05{\rm min}$, $\delta x=0.005{\rm cm}$.
We found numerically that for strong perturbations,
Eqn. (4-8) display a weak turbulence, Fig. 9.
Typically, for  $D_X \leq D_Y$, 
we found transient, weak turbulence. 
When  $D_X < < D_Y$, numerical experiments lead to
stationary turbulence. For instance, 
we simulated Eqn. (4-8)
up to  $T=10^7{\rm min}$ for $m=3$, 
$D_X=6\cdot10^{-7}\frac{ {\rm cm^2}}{{\rm min}}$, 
$D_Y=10^{-4}\frac{{\rm cm^2}}{{\rm min}}$, $D_Z=0$ 
and  $l=1.28{\rm cm}$ and found 
stationary turbulence for a number of different initial conditions.

\section{Discussion}
We have shown in this paper that two representative mathematical models of
biochemical oscillations exhibiting  birhythmicity, glycolytic and cell 
cycle models, display weak turbulence, intermiitency
of large and small amplitude oscillations. We revealed that 
unstable oscillations near cyclic fold bifurcations
are the mechanism of transient turbulence in birhythmic media.
In the presence of wavenumber instability, weak turbulence is stationary.

Recently,  Stich et. al. \cite{alex1,alex2} proposed an amplitude model for 
birhythmic media. An interesting question is whether the weak 
turbulence we discussed in this paper 
can be found in their model? First, let us mention two important 
differences between our models and the amplitude model of birhythmic media. 
In our case, a cyclic fold bifurcation is crucial for turbulence,
but the amplitude model describes a pitch-fork bifurcation
of limit cycles. Secondly, both fast and slow 
oscillations in the amplitude equation are smooth, but in our case, 
slow oscillations are strongly anharmonic. 
Besides these differences, it is well known that 
if phase slips develop,
the CGLE equation generates defects \cite{chate}.
Thus, these facts indicate that
instead of intermittency of small and large
amplitude oscillations, defect
 turbulence
is likely in the amplitude model of birhythmic media. 

To date, there are no experimental evidences of weak turbulence in glycolysis
or in the cell cycle. Our results are pure theoretical predictions of mathematical 
models. The system sizes we simulated are much larger 
than the typical size of an yeast cell ($10^{-3}{\rm cm}$). Therefore, weak 
turbulence is not expected in yeasts. Interestingly, 
some slime molds grow as syncytial 
plasmodia (many nuclei in a common cytoplasmic pool) that are many 
times larger than a typical yeast cell;
cells $15{\rm cm}$ in diameter can be grown in the laboratory \cite{moh}.
Waves of nuclear division are observed in these multinucleate
plasmodia \cite{haskin,tysonnovak}, and, as we have shown it is possible that 
these waves exhibit weak turbulence.  
Note that weak turbulence in the cell cycle would
mean irregular oscillations of CDK. But for a normal cell cycle, 
large amplitude oscillations of CDK are essential; CDK activity must drop below 
a certain threshold for nuclei to exit mitosis and divide. Therefore, 
hypothetically, weak turbulence in syncytial plasmodia might lead 
to mitotic arrest of certain nuclei in the plasmodium.

A more quantitative characterization of wavenumber instability of unstable 
oscillations at the onset of a cyclic fold bifurcation,
as well as simulations in two spatial 
dimensions \cite{bat} are problems in the future.

\appendix
\section{Coefficients of the  CGLE for a glycolytic model with substrate inhibition}
In this appendix, 
following standard procedures  in Ref. \cite{kura1}, 
we will calculate
coefficients of  CGLE for the glycolytic model.
For a convenience we assume $Q\equiv 1$ in Eqn. (1-3).
First, let us find uniform steady state solutions $\alpha_0$ and $\gamma_0$,
\begin{eqnarray}
\gamma_0=\mu/k_s, \hspace{6cm}\\
\alpha_0=\frac{
K^4 (-2\mu+\sigma)+\gamma_0^4(-2(\mu+\sigma_i)+\sigma) }
{\tilde c
}-\nonumber\hspace{2cm}\\
-\frac{
\sqrt{-4 \tilde a^2
+4 L\sigma \tilde a \tilde b
+(1+\gamma_0)^2\sigma^2\tilde b^2}
}
{
(1+\gamma_0)\tilde c
}\nonumber\hspace{2cm}\\
\end{eqnarray}
%%%%%%%%%%%%%%%%%%%%%%%%%%%%%%%%%%%%%%%%%%%%%%%%%%%%%%%%%%
%%%%%%%%%%%%%%%%%%%%%%%%%%%%%%%%%%%%%%%%%%%%%%%%%%%%%%%%%%
where, 
$\tilde a=K^4\mu+\gamma_0^4(\mu+\sigma_i)$,
$\tilde b=(K^4+\gamma_0^4)$,
$\tilde c=2(K^4 (\mu-\sigma)+\gamma_0^4(\mu+\sigma_i-\sigma))$.
%%%%%%%%%%%%%%%%%%%%%%%%%%%%%%%%%%%%%
Next we perform a linear stability analysis of $(\alpha_0,\gamma_0)$ 
against small fluctuations $\delta \alpha, \delta \gamma \propto
exp(iq x+i \lambda t)$. At the critical wavenumber $q_{cr}=0$, we obtain
a characteristic equation,
\begin{equation}
\lambda^2+(a_1+a_2+k_s)\lambda+a_1k_s=0.
\end{equation} 
In Eqn. (A3), 
$a_1$ and $a_2$ are given by 
\begin{equation}
a_1=
\frac{\sigma(1+\gamma_0)^2(L+2 L\alpha_0+(1+\alpha_0)^2(1+\gamma_0)^2)}
{(L+(1+\alpha_0)^2(1+\gamma_0)^2)^2},
\end{equation}
\begin{equation}
a_2=\frac{4 K^4 \gamma_0^3 \sigma_i}{(K^4+\gamma_0^4)^2}-
\frac{2\sigma L\alpha_0(1+\alpha_0)(1+\gamma_0)}
{(L+(1+\alpha_0)^2(1+\gamma_0)^2)^2}.
\end{equation} 
Let us define  such a critical value  for the bifurcation 
parameter $\sigma_i=\sigma_{i,cr}$  that 
\begin{equation}
a_1+a_2+k_s\equiv0.
\end{equation}
Eqn. (A6) is the condition for a Hopf bifurcation;
the characteristic equation has 
pure imaginary solutions,
$\lambda_0=\pm i\sqrt{a_1 k_s}$. 

Let $\mu$ be defined by $\mu=\frac{\sigma_i-\sigma_{i,cr}}{\sigma_{i,cr}}$. 
We develop the Jacobian matrix $L$ of Eqn. (1-3) in powers of $\mu$,
\begin{equation}
L=L_0+\mu L_1+....
\end{equation}
At $\mu=0$ the Jacobian  is given by
\begin{eqnarray}
L_0=
\left|
\begin{array}{c}
\begin{array}{cc}
-a_1\\
a_1
\end{array}
\begin{array}{cc}
\ \ \ \ a_2\\
\ \ \ \ -k_s-a_2.
\end{array}
\end{array}\right|.
\end{eqnarray} 
We find the right $\bf{u}_0$  and 
left ${\bf{u}_0}^*$ eigenvectors of $L_0$ corresponding to
$\lambda_0$, 
\begin{eqnarray}
 \bf{u}_0=
\left(
\begin{array}{c}
-1+ i \sqrt{\frac{k_s}{a_1}}\\
1
\end{array}\right),
\end{eqnarray} 
\begin{equation}
{ \bf{u}_0}^*= \frac{1}{2} \left(-i 
\sqrt{\frac{a_1}{ k_s}}, 1 - i \sqrt{\frac{a_1}{ k_s}}\right).
\end{equation} 
We find further,
\begin{eqnarray}
L_1=
\frac{4 K^4 \gamma_0^3  \sigma_{i,cr}}{(K^4+\gamma_0^4)^2 }\left|
\begin{array}{c}
\begin{array}{cc}
0\\
0
\end{array}
\begin{array}{cc}
\ \ \ \ -1\\
\ \ \ \ 1
\end{array}
\end{array}\right|.
\end{eqnarray} 
Let us first find $c_0$ in the CGLE.
It  is given by $c_0=\frac{Im\lambda_1}{Re\lambda_1}$, where
\begin{equation}
\lambda_1={ \bf{u}_0}^* L_1{\bf{u}_0}= 
\frac{2 K^4 \gamma_0^3}{(K^4+\gamma_0^4)^2} \sigma_{i,cr}.
\end{equation} 
We see that   $\lambda_1$ is pure real, therefore, 
$c_0=0$. Now following again \cite{kura1}, 
we find $c_1$, 
\begin{eqnarray}
D=
\left|
\begin{array}{c}
\begin{array}{cc}
D_{\alpha}\\
0
\end{array}
\begin{array}{cc}
\ \ \ \ 0\\
\ \ \ \ D_{\gamma}
\end{array}
\end{array}\right|,
\end{eqnarray} 
%%%%%%%%%%%%%%%%%%%%%%%%%%%%%%%%%%%%%%%%%%%%%%%%%%%%%%%%%%%%%%%%%%%%%%%%%%%%
%%%%%%%%%%%%%%%%%%%%%%%%%%%%%%%%%%%%%%%%%%%%%%%%%%%%%%%%%%%%%%%%%%%%%%%%%%%%

\begin{equation}
d'+id''={ \bf{u}_0}^* D \bf{u}_0,
\end{equation} 
%%%%%%%%%%%%%%%%%%%%%%%%%%%%%%%%%%%%%%%%%%%%%%%%%%%%%%%%%%%%%%%%%%%%%%%%%%%%
\begin{equation}
c_1=d''/d'=\sqrt{\frac{a_1}{k_s}}
\left(\frac{D_{\alpha}-D_{\gamma}}{D_{\alpha}+D_{\gamma}}\right).
\end{equation}
%%%%%%%%%%%%%%%%%%%%%%%%%%%%%%%%%%%%%%%%%%%%%%%%%%%%%%%%%%%%%%%%%%%%%%%%%%%%
Calculation of $c_2$ is a little more tedious. We need to find \cite{kura1}, 
\begin{eqnarray}
{\bf V}_+={\bf \bar{V}}_-=-{(L_0-2 \lambda_0)}^{-1}{\bf M}_0{\bf u}_0{\bf u}_0,\\
{\bf V}_0=-2 {L_0}^{-1} {\bf M}_0 {\bf u}_0 {\bar{\bf u}}_0,
\end{eqnarray}
%%%%%%%%%%%%%%%%%%%%%%%%%%%%%%%%%%%%%%%%%%%%%%%%%%%%%%%%%%%%%%%%%%%%%%%%%%%%
\begin{eqnarray}
g=g'+i g''= -2 { \bf{u}_0}^*{\bf M}_0{\bf u}_0 {\bf V}_0-
2  { \bf{u}_0}^*{\bf M}_0  {\bar{\bf u}}_0 {\bf V}\nonumber\\
-3 { \bf{u}_0}^* {\bf N}_0 { \bf{u}_0}^* { \bf{u}_0}^* {\bar{\bf u}}_0.
\hspace{1.5cm}
\end{eqnarray}
%%%%%%%%%%%%%%%%%%%%%%%%%%%%%%%%%%%%%%%%%%%%%%%%%%%%%%%%%%%%%%%%%%%%%%%%%%%%
Parameter $c_2$ in the CGLE is given by a formula, $c_2=g''/g'$. 
We find that $c_2=\frac{\tilde g''} {\tilde g'}$ , where
\begin{eqnarray}
\tilde g'=-3 k_s [k_s m_{ \alpha^2} (2 m_{\alpha^2}-m_{\alpha\gamma})
+a_1((2 m_{\alpha^2}-m_{\alpha\gamma})
(m_{\alpha^2}-m_{\alpha,\gamma}+m_{\gamma^2})\nonumber
\ \ \ \ \ \ \ \ \ \ \\
-k_s ({n_{\alpha^2\gamma}}-3 {n_{\alpha^3}}))
+3 a_1^2( {n_{\alpha\gamma^2}}-{n_{\alpha^2\gamma}}+ {n_{\alpha^3}}-
 {n_{\gamma^3}})], \ \ \ \ \ \ \ \ \ \ \ \ \ \ \ \ \ \\
%%%%%%%%%%%%%%%%%%%%%%%%%%%%%%%%%%%%%%%%%%%%%%%%%%%%%%%%%%%%%%%%%%%%%%%%%%%
\tilde g''=\sqrt{\frac{k_s}{a_1}}[10 k_s^2 {m_{\alpha^2}}^2+ 
a_1 k_s (14 {m_{\alpha^2}}^2-14 {m_{\alpha^2}}m_{\alpha,\gamma}
+{m_{\alpha\gamma}}^2+10 {m_{\alpha^2}} {m_{\gamma^2}}
+9k_s {n_{\alpha^3}})+\nonumber\\
+a_1^2 {(4 ({m_{\alpha^2}}-
m_{\alpha,\gamma}+{m_{\gamma^2}})}^2+
3 k_s ( {n_{\alpha\gamma^2}}-2 {n_{\alpha^2\gamma}}+3 {n_{\alpha^3}}))].
\ \ \ \ \ \ \ \ \ \ \ \ \ \ 
\end{eqnarray}
In the above expressions, 
$m_{\alpha^2}=(\frac{\partial^2 \phi(\alpha,\gamma)}
{\partial\alpha^2})_{\alpha_0,\gamma_0}$,
$m_{\alpha\gamma}=(\frac{\partial^2 \phi(\alpha,\gamma)}
{\partial\alpha\partial\gamma})_{\alpha_0,\gamma_0}$,
$m_{\gamma^2}=\frac{2 (3 K^8 \gamma_0^2-5 K^4 \gamma_0^6) \sigma_{i,cr}}
{(K^4+\gamma_0^4)^3}-
(\frac{\partial^2 \phi(\alpha,\gamma)}
{\partial\gamma^2})_{\alpha_0,\gamma_0}$, 
$ {n_{\alpha^3}}= 
(\frac{\partial^3 \phi(\alpha,\gamma)}
{\partial\alpha^3})_{\alpha_0,\gamma_0}$,
${n_{\alpha^2\gamma}}=
(\frac{\partial^3 \phi(\alpha,\gamma)}
{\partial\alpha^2\partial\gamma})_{\alpha_0,\gamma_0}$,
${n_{\alpha\gamma^2}}=
(\frac{\partial^3 \phi(\alpha,\gamma)}
{\partial\alpha\partial\gamma^2})_{\alpha_0,\gamma_0}$,
${n_{\gamma^3}}=
(\frac{\partial^3 \phi(\alpha,\gamma)}
{\partial\gamma^3})_{\alpha_0,\gamma_0}$. To save space we
do not present here cumbersome expressions for
these derivatives.

For parameters in Fig. 1 we find that $\sigma_{i,cr}=1.282$ and
$c_2\approx 2.21$. From Eqn. (A15) we see that if  $D_{\alpha}=
D_{\gamma}$, $c_1=0$. Therefore, $1+c_1 c_2>0$ for parameters used in
this paper. If $D_{\alpha}=0$, Eqn. (A15) gives the minimal
value, $c_1=-0.47$. In this case $1+c_1 c_2 \approx -0.03$, therefore,
wavenumber instability is possible. However, turbulence must be weak
as the parameters are very close to the stability condition $1+c_1 c_2>0$
\cite{cross}.  A stronger wavenumber instability is possible, for example, for 
$\mu=0.28{\rm s^{-1}}$,  $K=12$, 
$D_{\alpha}=5 \cdot 10^{-7}\frac{{\rm cm^2}}{{\rm s}} $, 
$D_{\gamma}=1\cdot 10^{-5} \frac{{\rm cm^2}}{{\rm s}}$
and other
papameters are the same as in Fig. 1. In this case, we find that
$\sigma_{i,cr}\approx 1.095$ and $1+c_1 c_2=-0.416$.

\listoffigures

FIG. 1.  A bifurcation diagram of Eqn. (1-3). {\rm HB} marks 
a Hopf bifurcation point, ${\rm CF^{1,2}}$  mark cyclic fold
bifurcations. Parameters are:
$\nu =0.25{\rm s^{-1}}$, $n=4$, $K=11.5$, $\sigma=11{\rm s^{-1}}$, 
$k_s=0.05{\rm s^{-1}}
$, $L=3400000$.

FIG. 2.  Space-time pattern of $\gamma$ in 
a weak turbulent regime of Eqn. (1-2).
The space and time spans are $l=1.75{\rm cm}$ 
and $T=5\cdot10^3{\rm s}$. The pattern was obtained
by recording $\gamma(x)$ with a time interval $\tau=5{\rm s}$.
$D_{\alpha}=D_{\gamma}=1\cdot 10^{-5} \frac{{\rm cm^2}}{{\rm s}}$
and $\sigma_i=1.065{\rm s^{-1}}$.  
Other parameters are the same as in Fig. 1.

FIG. 3.  Snapshots of spatial distributions of $\alpha$ at two different
time moments. Parameters are the same as in Fig. 2.

FIG. 4.  A phase plane view. The outer cycle shows the 
orbit of stable uniform oscillations with a period $\tau=300{\rm s}$.
The inner cycle shows the orbit of small amplitude,
fast oscillations with a period  $\tau=290 {\rm s}$ at 
$\sigma_i=1.08{\rm s^{-1}}$.
With the decrease of $\sigma_i$, the inner cycle disappears,
but it still can  attract neighboring trajectories creating
a virtual, chaotic heterogeneity in Eqn. (1-2). The solid lines 
show oscillator distributions at two different time moments.
Parameters are the same as in Fig. 2.

FIG. 5.  A log-log plot of the 
spatial correlation function. Parameters are the same as in Fig. 2.

FIG. 6   Average transient lifetime versus the system size. The inset
shows the cases when a collapse of turbulence has not occurred by
$T=10^6{\rm s}$. Parameters are the same as in Fig. 2.

FIG. 7.  Breathing periodic structures. $l=3.5 {\rm cm}$,
other parameters, as well as the time 
and space spans are the same as in Fig. 2. 

FIG. 8.  Bifurcation diagram of a cell cycle model.
Rate constants  $k_i$ are in units ${\rm min^{-1}}$,
$k_1=0.002$, $k_2=0.053$,
$k_3=0.01$, $k_4=2$, 
$k_5=0.05$, $k_6=0.04$,
$k_7=1.5$, $k_8=0.19$, $k_9=0.64$, 
$k_{10}=0.005$, $k_{11}=0.07$, $k_{12}=0.08$.
Other parameters are 
$P=0.15$, $J_1=0.05$, and $J_2=0.01$, 
$l=1.28{\rm cm}$, 
$D_X=6\cdot10^{-7}\frac{{\rm cm^2}}{{\rm min}}$, 
$D_Y=10^{-4}\frac{{\rm cm^2}}{{\rm min}}$ and $D_Z=0$.

FIG. 9. Turbulence in a cell cycle model.
Space time plot of $Y$ field in Eqn. (4-8). 
The space and time spans are $L=1.28{\rm cm}$ 
and $T=2500{\rm min}$. The pattern was obtained
by recording $Y(x)$ with a time interval $\tau=5{\rm min}$.  

%\newpage
%\newpage
%\centerline{
%\epsfig{file=Figure1.eps,scale=1,angle=0}}
%\bigskip
%\centerline{{\LARGE Figure 1}}

%\newpage
%\bigskip
%\centerline{{\LARGE Figure 2}}
%\epsfig{file=Figure2.eps,scale=1,angle=0,
%bbllx=100,bblly=500,bburx=350,bbury=700}

%\newpage
%\centerline{
%\epsfig{file=Figure3.eps,scale=1,angle=0}}
%\bigskip
%\centerline{{\LARGE Figure 3}}

%\newpage
%\centerline{
%\epsfig{file=Figure4.eps,scale=1,angle=0}}
%\bigskip
%\centerline{{\LARGE Figure 4}}

%\newpage
%\centerline{
%\epsfig{file=Figure5.eps,scale=1,angle=0}}
%\bigskip
%\centerline{{\LARGE Figure 5}}

%\newpage
%\centerline{
%\epsfig{file=Figure6.eps,scale=1,angle=0}}
%\bigskip
%\centerline{{\LARGE Figure 6}}

%\newpage
%\bigskip
%\centerline{{\LARGE Figure 7}}
%\epsfig{file=Figure7.eps,scale=1,angle=0,
%bbllx=100,bblly=500,bburx=350,bbury=700}

%\newpage
%\centerline{
%\epsfig{file=Figure8.eps,scale=1,angle=0}}
%\bigskip
%\centerline{{\LARGE Figure 8}}

%\newpage
%\bigskip
%\centerline{{\LARGE Figure 9}}
%\epsfig{file=Figure9.eps,scale=1,angle=0,
%bbllx=100,bblly=500,bburx=350,bbury=700}


\begin{thebibliography}{}
\bibitem{bat} Our preliminary results  show existence of the weak 
turbulence in dimension two.
\end{thebibliography}
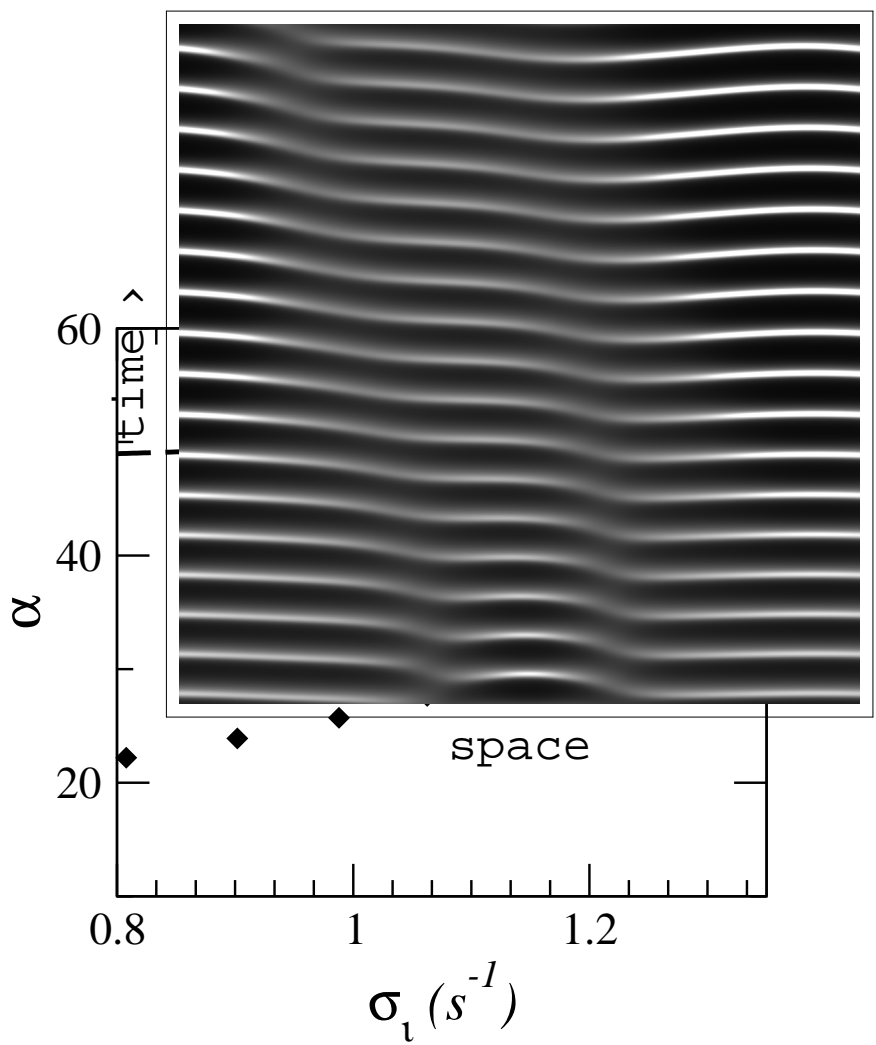
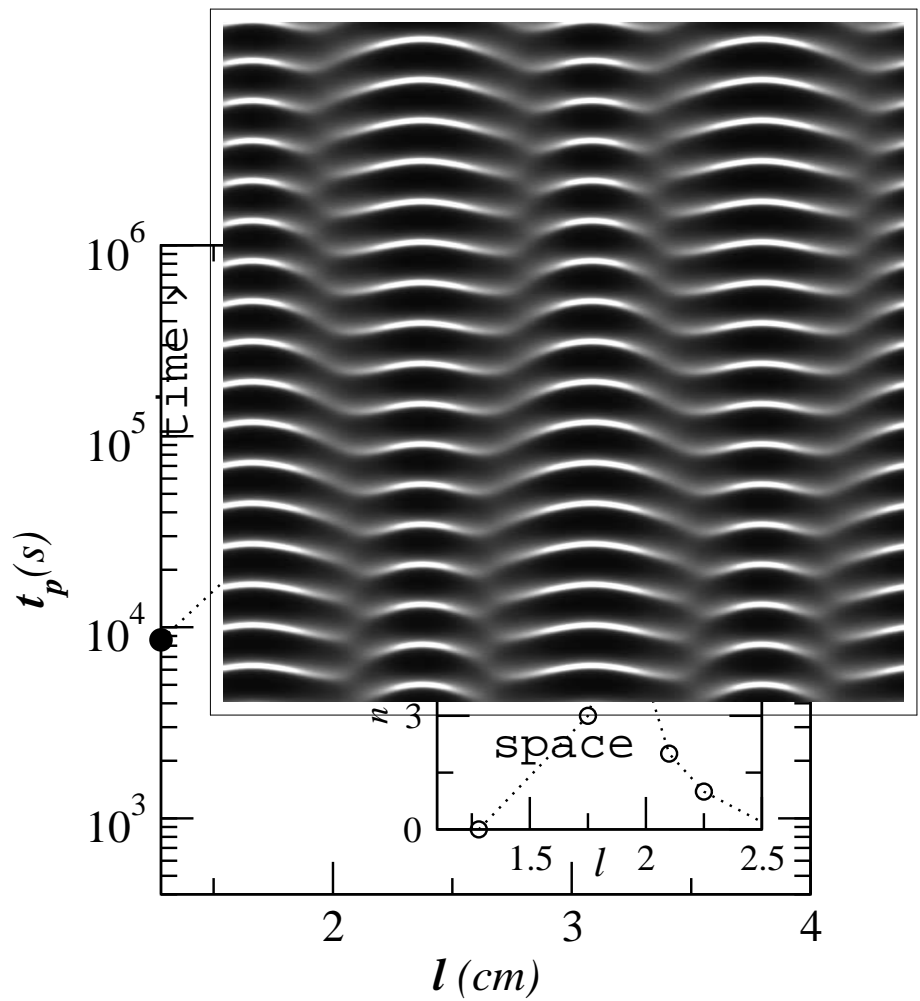
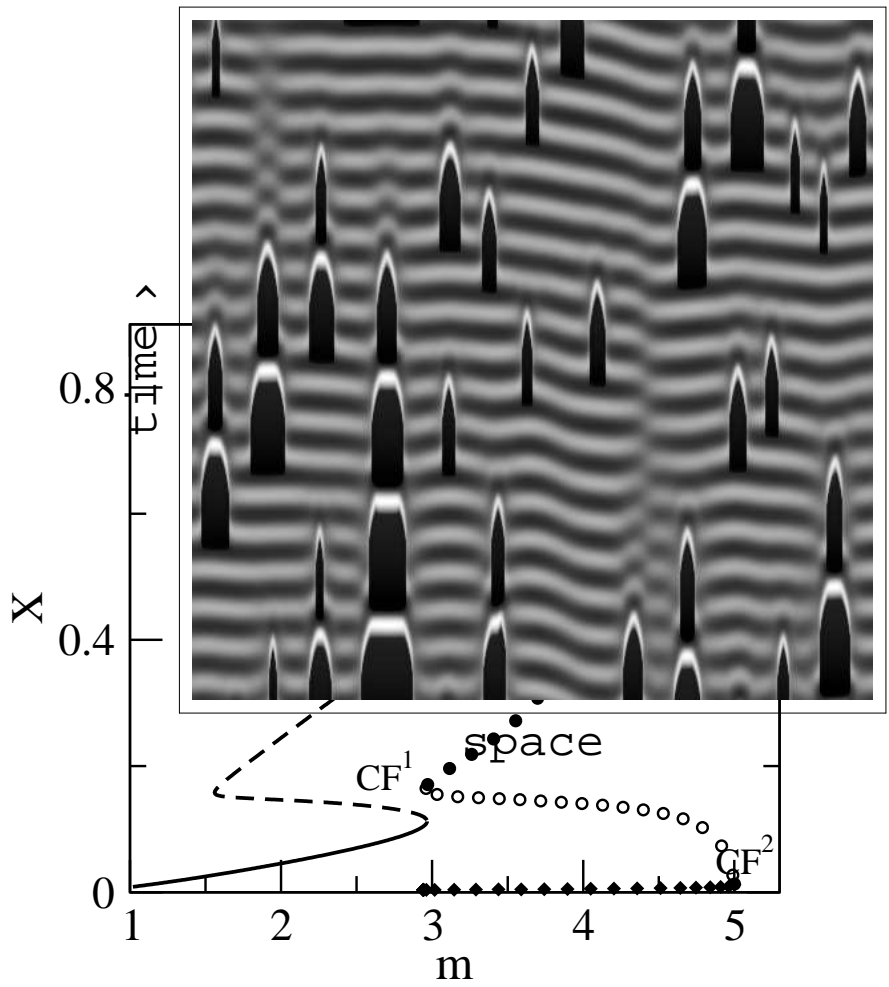
\end{document}